\documentclass[letterpaper, 10 pt, conference]{ieeeconf}  
\IEEEoverridecommandlockouts  
\usepackage{graphicx}
\usepackage{indentfirst}
\usepackage{array,multirow}
\usepackage{fancyhdr}
\usepackage{graphicx}
\usepackage{caption}
\usepackage{float}
\usepackage{booktabs}
\usepackage{array,multirow}
\usepackage{times} 
\usepackage{here}
\usepackage{comment}
\usepackage{CJKutf8}
\usepackage{hyperref}
\usepackage{fontawesome5}
\usepackage{color}
\IEEEaftertitletext{\vspace{-0.9\baselineskip}}

\newcommand{\ctext}[1]{\raise0.2ex\hbox{\textcircled{\scriptsize{#1}}}}

\title{\LARGE \bf
Importance of Instruction for Pedestrian--Automated Driving Vehicle Interaction with an External Human Machine Interface:
\\ \vspace{-1mm}{\large Effects on Pedestrians' Situation Awareness, Trust, Perceived Risks and Decision Making}\vspace{-3mm}
}

\author{Hailong Liu~$^{1}$,~\IEEEmembership{Member,~IEEE}, Takatsugu Hirayama~$^{2}$,~\IEEEmembership{Member,~IEEE} and Masaya Watanabe~$^{3}$
\thanks{$^{1}$~Hailong Liu is with Graduate School of Informatics, Nagoya University, Furo-cho, Chikusa-ku, Nagoya, Aichi, 464-8601, JAPAN.
 \faIcon{envelope}~:~{\tt\small lhl881210@live.com} (corresponding author)~\textcolor{white}{\faIcon{pastafarianism}}} 
\thanks{$^{2}$~Takatsugu Hirayama is with University of Human Environments, 6-2, Kamisanbonmatsu, Motojuku-cho, Okazaki, Aichi, 444-3505, JAPAN.
 \faIcon[regular]{envelope}~:~{\tt\small t-hirayama@uhe.ac.jp}} 
\thanks{$^{3}$~Masaya Watanabe is with Vehicle Development Center,
Toyota Motor Corporation, Toyota-cho, Toyota, Aichi, 471-8572, JAPAN.
\faIcon[regular]{envelope}~:~{\tt\small masaya\_watanabe@mail.toyota.co.jp}}%
}

\begin{document}
\thispagestyle{empty}
\pagestyle{empty}
\maketitle
\begin{abstract}
Compared to a manual driving vehicle~(MV), an automated driving vehicle lacks a way to communicate with the pedestrian through the driver when it interacts with the pedestrian because the driver usually does not participate in driving tasks.
Thus, an external human machine interface~(eHMI) can be viewed as a novel explicit communication method for providing driving intentions of an automated driving vehicle~(AV) to pedestrians when they need to negotiate in an interaction, e.g., an encountering scene.
However, the eHMI may not guarantee that the pedestrians will fully recognize the intention of the AV.
In this paper, we propose that the instruction of the eHMI's rationale can help pedestrians correctly understand the driving intentions and predict the behavior of the AV, and thus their subjective feelings (i.\,e., dangerous feeling, trust in the AV, and feeling of relief) and decision-making are also improved.
The results of an interaction experiment in a road-crossing scene indicate that the participants were more difficult to be aware of the situation when they encountered an AV w/o eHMI compared to when they encountered an MV; further, the participants' subjective feelings and hesitation in decision-making also deteriorated significantly. 
When the eHMI was used in the AV, the situational awareness, subjective feelings and decision-making of the participants regarding the AV w/ eHMI were improved.
After the instruction, it was easier for the participants to understand the driving intention and predict driving behavior of the AV w/ eHMI. Further, the subjective feelings and the hesitation related to decision-making were improved and reached the same standards as that for the MV.
\end{abstract}

\section{INTRODUCTION}
Automated driving vehicles~(AVs) are expected to be widely used in the near future~\cite{Vissers2016,rasouli2019}.
Under this assumption, some traffic scenarios are expected to contain a mixture of AVs and pedestrians such as shared spaces, intersections with no traffic lights, narrow roads, and parking lots.
However, compared to manual driving vehicles~(MVs), AVs lack a way to directly communicate with pedestrians.
This may cause problems such as AVs being difficult to share driving intentions and negotiate the right-of-way with pedestrians.

In the interaction between MVs and pedestrians, the latter often understand the intention of the MVs through implicit information such as the speed, acceleration, and steering angular velocity of the vehicle, and the direction of the wheels~\cite{rasouli2019}.
Further, communication through explicit information, e.g., the head nod and hand gestures, can help pedestrians understand the intention of the driver and resolve ambiguity of negotiation in interactions~\cite{sucha2017pedestrian,farber2016communication}.
A typical example is when a pedestrian encounters an MV on a narrow road without traffic signals or in a parking lot; in this case, the driver can clearly convey their intention and quickly reach an agreement of a  negotiation with the pedestrians via eye contact, hand gestures, and verbal communication. 

However, for AVs driven by levels 3--5 automated systems, the driver does not participate in the driving task~\cite{SAE_j3016_2016}.
Therefore, the AV may lack a useful explicit communication in its interaction with pedestrians, particularly in complex urban environments and shared spaces where explicit communication methods such as hand gestures are frequently used between MV's drivers and pedestrians today~\cite{Vissers2016}.
This lack makes pedestrians difficult to understand the intentions of the AV quickly and clearly~\cite{liu2020what,liu2020when}.
It also causes potential issues, such as safety hazards, inefficiency, and poor pro-sociality~\cite{batson2003altruism}.
Thus, understanding how the AVs should communicate with pedestrians has become a pressing concern.

To solve this issue, a novel communication approach using an external human machine interface~(eHMI) can be viewed as one of the solutions~\cite{schieben2019designing}. In particular, various studies have evaluated the efficacy of eHMIs for presenting the intention of an AV to pedestrians using light bars, icons, and text~\cite{rettenmaier2019passing,Stefanie2020}.
Although these studies advocate good eHMI designs, these eHMIs cannot guarantee that the pedestrians will fully recognize the intention of the AV, fully understand the rationale behind the AV's intention and perceive limitations of AV's functional abilities, especially those who do not have considerable experience with eHMIs.

\begin{figure*}[h!tb]
\centering
\includegraphics[width=0.95\linewidth]{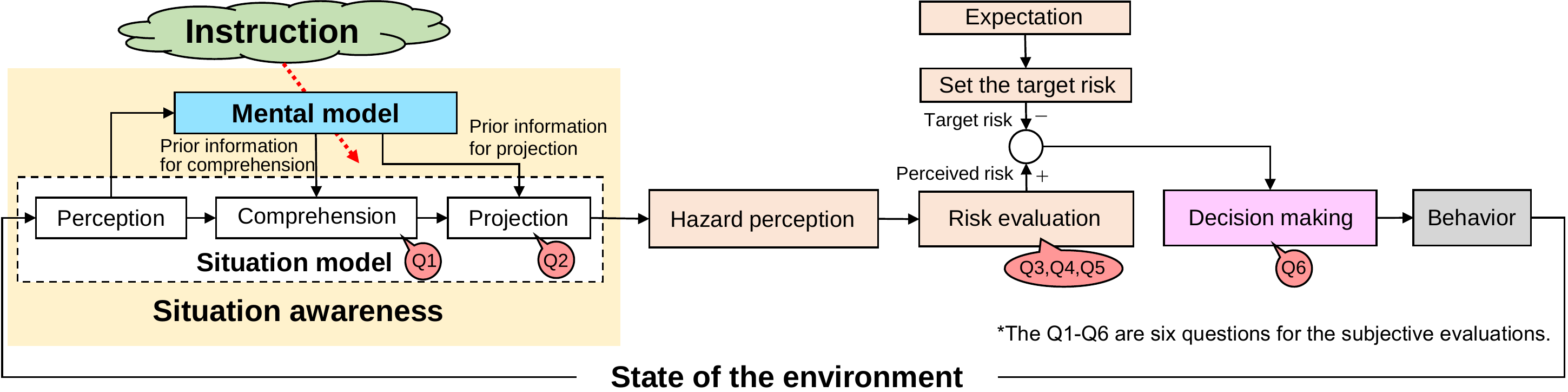}
\vspace{-1mm}
\caption{The proposed cognitive-decision-behavior model of a pedestrian based on the model in~\cite{liu2020when}.}
\label{fig:model}
\vspace{-6mm}
\end{figure*}

For example, even if the pedestrians clearly recognize that the AV has yielded the right of way via the message ``you go first'' displayed on the eHMI, it may be difficult for the pedestrians to understand under what circumstances the AV will display this information, especially when the sensor range is unknown. Furthermore, the pedestrians may be unsure about the time available for them to cross the road when the message is displayed because it may be difficult for the pedestrians to predict when the vehicle will depart.

To solve the above problem of the eHMI, one solution is to increase the number of interactions between pedestrians and AVs equipped with eHMI to ensure that they learn the intention of the AV through the information on eHMI~\cite{Michal2020}.
However, this method requires pedestrians to pay the time cost of learning. 
Further, pedestrians may remain in danger because of insufficiently understanding the intention of the AV during the learning process.
We consider that the fundamental approach to solve this problem is to help pedestrians establish the correct understanding quickly using a mental model~\cite{endsley1995toward} of the AV with eHMI. 
In this study, we propose a simpler approach that instructs pedestrians to understand the execution conditions, mechanisms, and function limitations of the AV and the eHMI. 
Further, this approach helps pedestrians gain situational awareness for interacting with the AV and to understand the intention and predict the behavior of the AV correctly.
Thus, this approach can help pedestrians improve the subjective feelings regarding AVs, which can increase the social acceptance of AVs.

Many related studies have reported that instructions for an in-vehicle human machine interface~(iHMI) of the AV have enabled drivers improve their understanding of using the AV, interactive performance with the AV, and trust in the AV~\cite{hergeth2017prior, forster2019user, edelmann2020effects}. However, owing to the growing popularity of AVs, both drivers and pedestrians interact with AVs, and therefore, instructing pedestrians to understand the intention of the AV correctly when interacting with it is an urgent issue that is yet to be studied widely.

In this paper, we propose using an instruction to help pedestrians establish a correct mental model of eHMI on AV.
We investigate changes in situational awareness and various subjective evaluations of pedestrians when interacting with vehicles under various scenarios based on a cognitive-decision-behavior model of pedestrians. 
The experimental results not only verify the effectiveness of eHMI for the interaction of pedestrians and AVs, but also that the instruction of the rationale of eHMI can improve the situational awareness, subjective feelings and decision-making of the pedestrians towards AVs.

\section{PURPOSE AND HYPOTHESIS}\label{sec:Hypercese}

In this study, we aim to analyze the influence of the instructing the rationale of eHMIs on the situational awareness, subjective feelings and decision-making of pedestrians when they encounter an AV. We verify the following hypothesis:

\begin{enumerate}
\item[H:] If pedestrians correctly understand the rationale of eHMI through the instruction, then the eHMI on AV can help them to improve the situational awareness, subjective feelings and decision-making during the interaction.
\end{enumerate}

To verify this hypothesis, we further consider the mechanism of the mental model in the cognitive-decision-behavior process of pedestrians based on a model proposed in our previous studies~\cite{liu2020what,liu2020when}, which is shown in Fig.~\ref{fig:model}. This cognitive-decision-behavior model includes three parts: situational awareness, risk evaluation based on hazard perception, and decision making based on risk homeostasis. The situational awareness of a pedestrian includes their ability to perceive objects in the surrounding environment (i.\,e., perception), understand its state and intention (i.\,e., comprehension), and predict its state in the future (i.\,e., projection). Then, the pedestrian perceives hazards based on their prediction results and evaluates the magnitude of the subjective risk. Next, the pedestrian decides the behavior by comparing the subjective risk with the acceptable risk level. This risk compensation process can be explained by the risk homeostasis theory~\cite{wilde1982theory}.

The mental model is an internal representation that contains meaningful declarative and procedural knowledge generated from long-term experiences~\cite{Al-Diban2012}. It is spontaneously generated by recognizing and interpreting the system by repeatedly using it~\cite{staggers1993mental}. Further, the situation model is the current instantiation of the mental model, that is, the relationship between the mental model and the situation models is a two-layer structure~\cite{endsley2000}. The situation model can be viewed as a prediction model constructed and supported by an underlying mental model~\cite{endsley1995toward,mogford1997mental}. We consider that the mental model provides the situational model with some prior information and knowledge to guide it to perform comprehension and projection for a given situation.

Based on the above arguments, we propose forming and calibrating a mental model correctly by instructing pedestrians with relevant knowledge about the eHMI for interacting with the AV. We design an experiment to verify the effectiveness of the eHMI for the pedestrian-AV interaction; further, we demonstrate the significance of the instruction in improving the situational awareness, subjective feelings and decision-making of pedestrians during the interaction.

\begin{figure*}[tb]
\centering
\begin{minipage}[t]{0.49\textwidth}
\centering
\includegraphics[width=0.90\linewidth]{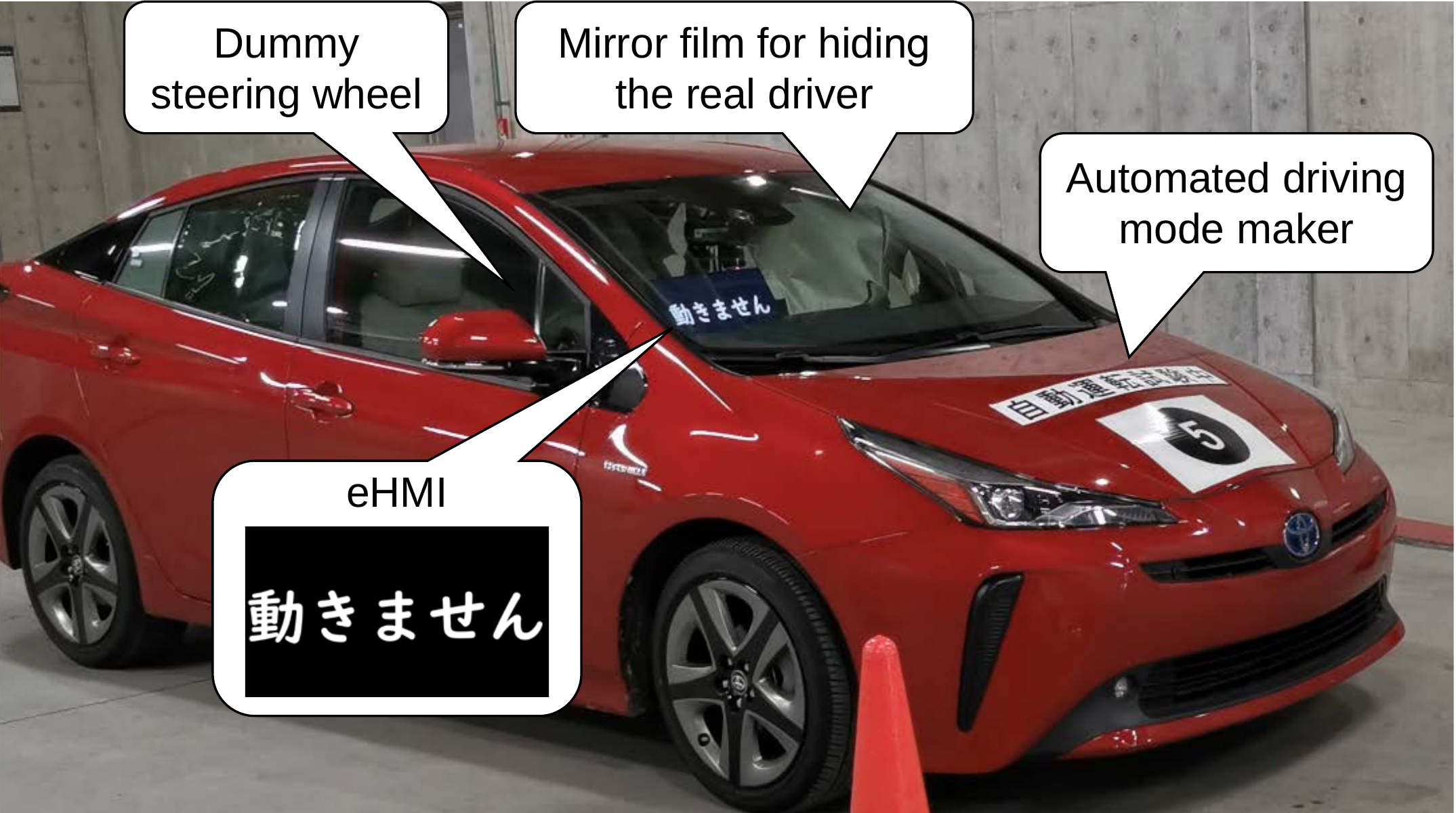}
\caption{Experimental vehicle: a left-hand driven car is used to simulate a right-hand driven AV.}
\label{fig:exp_car}
\end{minipage}
\hspace{1mm}
\begin{minipage}[t]{0.49\textwidth}
\centering
\includegraphics[width=0.90\linewidth]{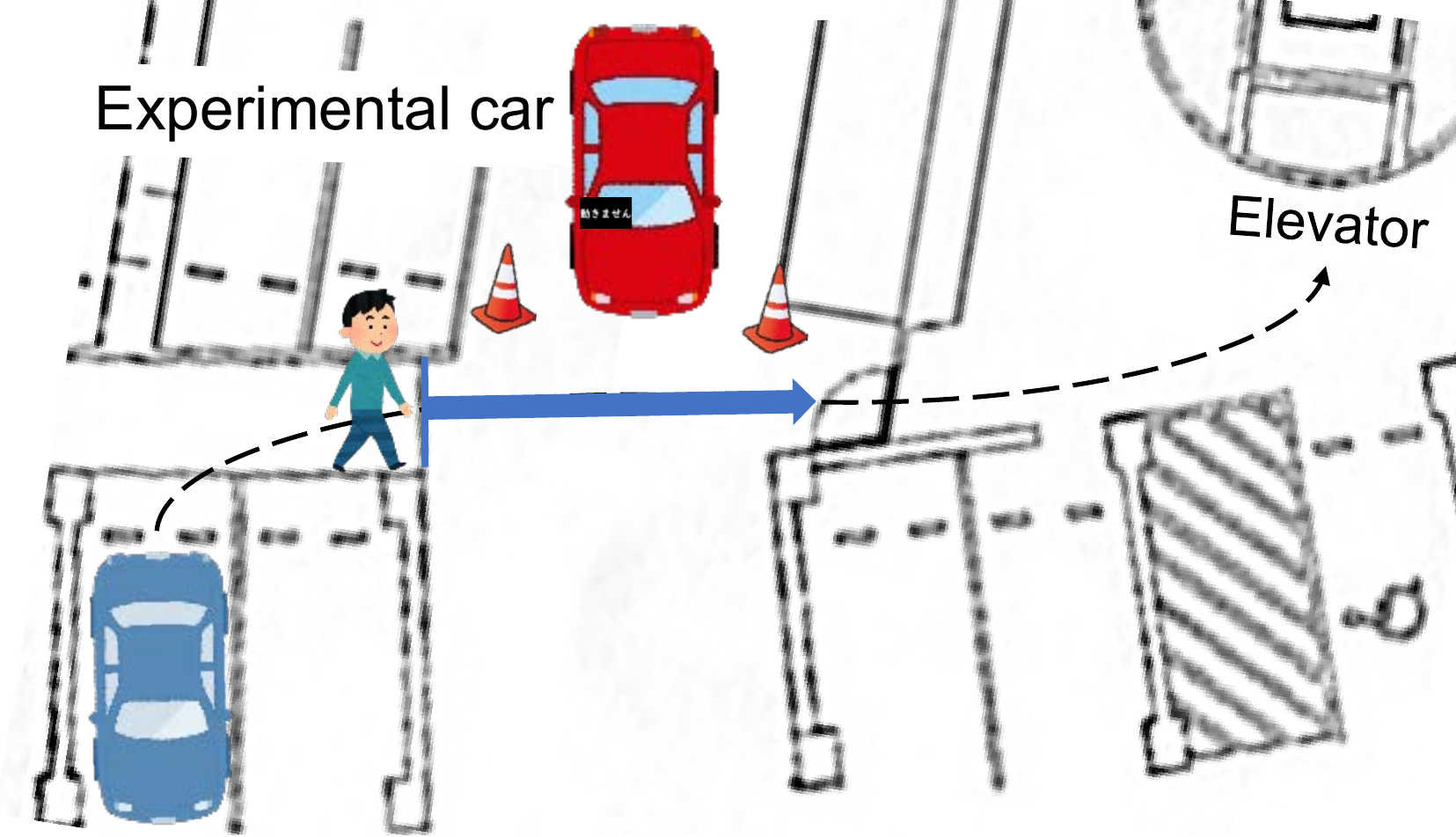}
\caption{Experimental scene: simulating a pedestrian encounters with a car in a parking lot.}
\label{fig:map}
 \end{minipage}
\vspace{-6mm}
\end{figure*}

\section{WIZARD OF OZ EXPERIMENT}\label{sec:experiment}
A pedestrian--vehicle interaction experiment was conducted in the B2F parking lot of the Toyota Stadium, Toyota-shi, Aichi, Japan. The safety of the experiment site (i.\,e., the B2F parking lot) was ensured by blocking access to the public. This experiment was approved by the ethics review committee of the Institute of Innovation for Future Society, Nagoya University.

\subsection{Experimental vehicle}
A left-hand-drive car (TOYOTA Prius) was used as the experimental vehicle which is shown in Fig.~\ref{fig:exp_car}. 
In general, cars in Japan are right-hand-drive cars. 
Therefore, to simulate an unmanned driving vehicle, i.e. AV, a real driver who is an expert driver was hidden behind on the left seat of the left-hand-drive car by using a mirror film.
Besides, a dummy steering wheel was installed on the right side to make the participants believe that it is a right-hand-drive car.
Considering the experimental scene of parking lot and the safety of participants, the maximum speed of the car was restricted to 8~km/h and the average speed was about 4~km/h.

\subsection{Design of eHMI}\label{eHMI}
The eHMI device is set behind the right side of the windshield.
After the AV stops, a message ``\begin{CJK}{UTF8}{ipxm}動きません\end{CJK}'' (``\textit{UGOKIMASEN}'') is displayed immediately. This message implies that the car does not move now.
The eHMI blinks twice in two seconds intervals after the pedestrian has crossed the road. This indicates that the car will move. After blinking, the eHMI is turned off and the car departs.

There are two reasons for considering the above settings:
(1) We do not want the car to command pedestrians because of liability. ``\textit{UGOKIMASEN}'' only indicates the current state and driving intention of the car to the pedestrians. It does not inform the pedestrians on what to do such as ``You go first.'' 
The pedestrians need to decide their walking behavior by themselves based on the state of the vehicle and the message on the eHMI. 
(2) This message can help us compare the effectiveness of the instruction because it can help pedestrians gain a vague understanding of the AV's intention in the specific context of Japanese. 
For example, the pedestrians may think that the AV is asking for help because the car has broken down when ``\textit{UGOKIMASEN}'' is displayed after the car halts. Moreover, the timing and action conditions of the message blinking are also unclear for the pedestrians. This is an important point that the pedestrians should be instructed about.

\subsection{Pedestrians}
32 participants participated in this experiment as pedestrians.
They are within the age range of 23--68 years (mean: 49.12, standard deviation: 11.13). Further, 17 participants are females and the remaining 15 were males. Before the experiment, the following information was provided to the participants:
\begin{enumerate}
\item Imagine you drive to the shopping center. You park your car in the underground parking lot and want to go to the elevator.
\item You need to cross a road to get to the elevator. Please walk at a normal speed during this process (see the dotted line in Fig.~\ref{fig:map}).
\item When you cross the road, a manual driving car, i.e. MV, or an automated driving car, i.e. AV, will arrive. You should be mindful of it when crossing the road.
\item The AV is a driver-less car. It is equipped with advanced built-in sensors that can detect pedestrians and the surrounding environment such as roads and the stop line. \textbf{(False information)}
\item For both the MV and AV, the pylons indicate a stop line (see Fig.~\ref{fig:map}). The car will stop before the stop line. After stopping, the car will decide whether to depart from the surrounding situation based on whether there is a pedestrian.
\end{enumerate}
It is important to note that information about the eHMI was not provided to the participants in this introduction before the experiment.

\subsection{Interaction scenarios}
\begin{figure*}[tb]
\centering
\includegraphics[width=0.86\linewidth]{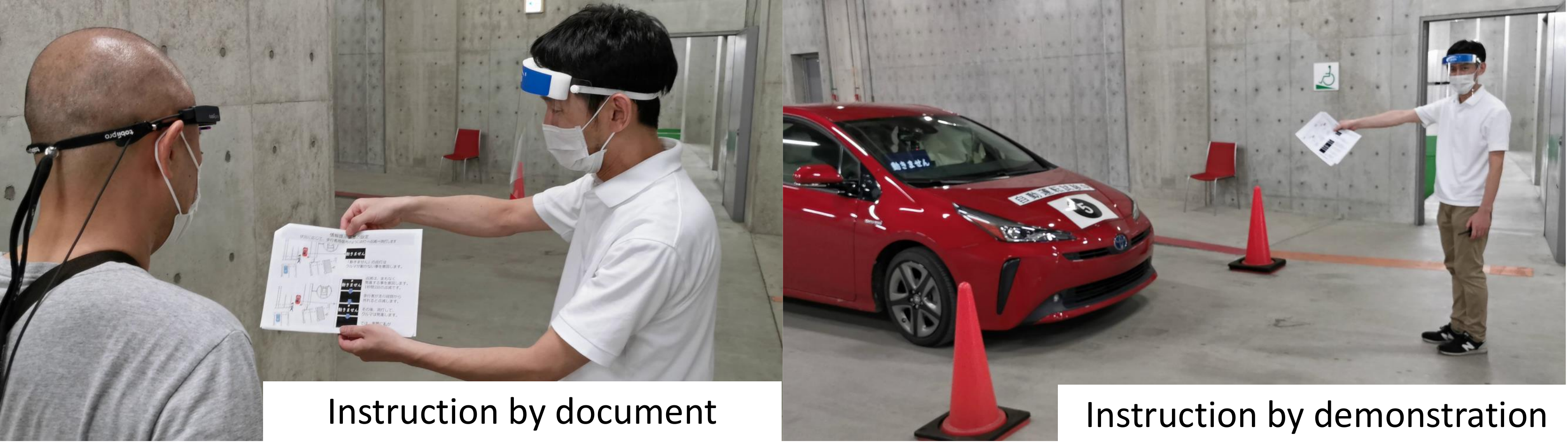}
\vspace{-1mm}
\caption{Scenes of instructing the rationale of eHMI on AV to the participant.}
\label{fig:instruction}
\vspace{-6mm}
\end{figure*}

Four scenarios were designed to allow the pedestrians to interact with the car in the following order: \textbf{MV}, \textbf{AV w/o eHMI}, \textbf{AV w/ eHMI} and \textbf{AV w/ eHMI after instruction}.
This order is to imitate the popularization order of the MV and the AV as well as the eHMI.
Each scenario was executed five times for each pedestrian.
In total, each participant encountered 20 vehicle trials.
The details of the four scenarios are the following.

\textbf{MV:} In this scenario, a pedestrian encounters an MV. A dummy driver sits on the right seat and imitates a real driver holding the steering wheel to control the car. When the pedestrian encounters the MV, the pedestrian can see the dummy driver driving the car.
Further, the dummy driver uses a typical Japanese gesture of "Please, go ahead", i.e., moving one's hand forward only once with his/her palm up (looks like \faIcon{hand-holding}), to indicate to the pedestrians to cross the road after stopping the car.

\textbf{AV w/o eHMI:} This is a scenario wherein the pedestrian encounters an AV. There is no eHMI device on the car; however, a striking marker is present on the hood of the car, which indicates that the car is in the automated driving mode (see Fig.~\ref{fig:exp_car}). Further, there is no dummy driver sitting on the right seat. When the car encounters the participant, the car stops before the stop line (two pylons). At this time, the participant needs to decide the timing of crossing the road and their walking behavior. Then, the car departs after the participant completely crosses the road.

\textbf{AV w/ eHMI:} This is a scenario wherein the pedestrian encounters an AV with an eHMI. The eHMI device is installed behind the right side of the windshield. The car stops in front of the stop line, and then, the eHMI shows the message ``\textit{UGOKIMASEN}'' to let the participant know that the car will not move. the participant needs to decide the timing for crossing the road and their walking behavior.
After the participant completely crosses the road, the message on the eHMI blinks twice. Then, the eHMI is turned off and the car departs.

\textbf{AV w/ eHMI after instruction:} This scenario is the same as the one before, i.\,e., AV w/ eHMI. However, the difference is that before the participant interacts with the vehicle, the pedestrian is instructed to create a correct mental model of the AV w/ eHMI.
Thus, the following information about the usage conditions and meaning of the eHMI are provided to the participant:
\begin{enumerate}
\item When the AV detects the pedestrian, the message ``\textit{UGOKIMASEN}'' will be displayed on the eHMI after the car stops. This indicates that the AV will not move. 
\item The eHMI will blink twice in two seconds after the pedestrian crosses the road. This indicates that the AV will move again and depart.
\item After blinking, the eHMI is turned off. Then, the AV will depart.
\end{enumerate}
This instructional scene is shown in Fig.~\ref{fig:instruction}.
To eliminate the vague understanding of the participant regarding the information on the eHMI, a document containing the above information was used to explain the meaning of the information on the eHMI. Further, a demonstration was used to explain to the participants when the eHMI would be turned on and when the eHMI would blink. The participants stood on the side of the road to watch the demonstrator's explanation. 
After the instruction, we answered questions raised by the participants about the content of the instruction. We conducted the experiments with the fourth scenario after confirming that the participants understood the content of the instruction.
We considered that the instruction helped them build a correct mental model of the AV w/ eHMI.

\subsection{Subjective evaluations for four scenarios}

After each trial, the participants were asked to use a five-grade evaluation scale---``Strongly Agree,'' ``Agree,'' ``Undecided,'' ``Disagree,'' and ``Strongly Disagree''---to respond to the following questions:
\begin{enumerate}
\item[Q1:] Was it easy to understand the driving intention of the car?
\item[Q2:] Was it easy to predict the behavior of the car?
\item[Q3:] Did you feel the behavior of the car was dangerous?
\item[Q4:] Did you trust the car when you crossed the road?
\item[Q5:] Did you feel a sense of relief when you crossed the road?
\item[Q6:] Did you hesitate when you crossed the road?
\end{enumerate}
As shown in Fig.~\ref{fig:model}, Q1 and Q2 are used to evaluate the comprehension and projection steps in the situation model; Q3, Q4, and Q5 are used for risk evaluation; and Q6 is used for evaluating the hesitation to make decisions, i.e., the difficulty of the decision-making for crossing the road.

\section{RESULTS AND DISCUSSIONS}

\begin{figure*}[tb]
\centering
\includegraphics[width=1\linewidth]{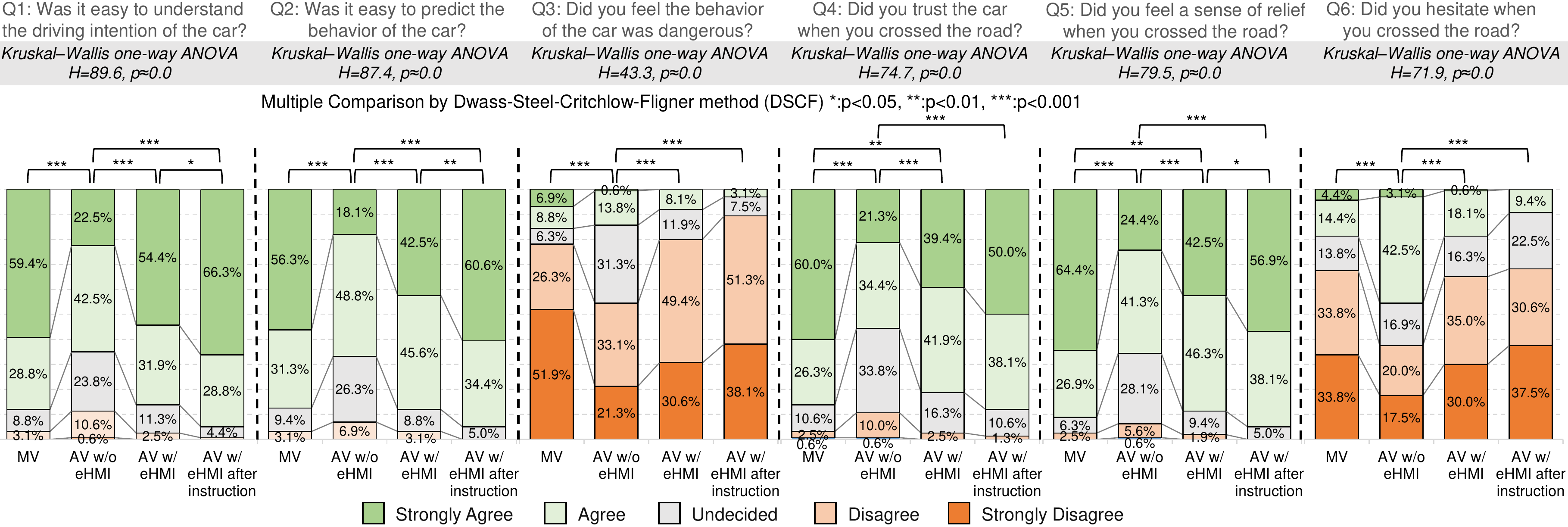}
\vspace{-6mm}
\caption{Results of subjective evaluations for four scenarios. (The ratios are rounded to the first decimal place)}
\label{fig:results}
\vspace{-6mm}
\end{figure*}

To verify that instructing the pedestrians about the eHMI helps them correctly understand the driving intentions and predict the behavior of the AV to improve their subjective feelings about the AV (i.\,e., the dangerous feeling, trust in the AV, and relieved feeling when they interact with the AV) and the decision-making, we analyzed the subjective evaluations of the participants for each scenario using six questions.
The subjective evaluations for each scenario were collected after each trial. For each question, the subjective evaluations of the participants under different scenarios were counted separately. Figure~\ref{fig:results} shows the proportions of the five-grade evaluation for the six questions for each scenario. The subjective evaluations of the pedestrians encountering the MV are used as the baseline. We compare the evaluation results from the other three scenarios with the baseline to discuss the problem of pedestrian-AV interaction, and we illustrate the effectiveness of our proposed solution, i.\,e., using eHMI with instruction. Multiple differences were observed in the results of the six questions for the four scenarios when compared using the Dwass--Steel--Critchlow--Fligner (DSCF) method~\cite{douglas1991distribution}.

\subsection{Evaluation of situational awareness based on Q1 and Q2}

According to the cognitive-decision-behavior model shown in Fig.~\ref{fig:model}. Q1 and Q2 are used to evaluate the comprehension and projection steps in the situation model of the participants. These results present the degree of difficulty for the participants to understand the driving intention and predict the behavior of the car.
Thus, the results of Q1 and Q2 can be used to demonstrate the effect of the instruction on the mental model because the situation model is affected by the mental model.

As the baseline, the results of Q1 in Fig.~\ref{fig:results} show that the participants in 59.4\% of trials agreed strongly that the driving intention of the MV was easy to understand. Only 3.1\% of trials were evaluated by Q1 as ``Disagree'' for the MV. However, the results of using the DSCF showed that there were significant differences between the subjective evaluations of participants encountering the AV w/o eHMI and the MV using Q1 (p$<$0.001) and Q2 (p$<$0.001). The driving intention of the AV became more difficult to understand; 22.5\% and 10.6\% of trials were evaluated as ``Strongly Agree'' and ``Disagree'' for Q1, respectively. Meanwhile, the results of Q2 indicate that the behavior of the MV was easier to predict than the behavior of the AV w/o eHMI because 56.3\% of trials wherein the participants encountered the MV were evaluated as ``Strongly Agree'' for Q2; however, only 18.1\% of trials wherein the participants encountered the AV w/o eHMI were evaluated as ``Strongly Agree.'' This could be because the participants did not have a way to understand the driving intention without receiving explicit information from the AV, and therefore, it became difficult for them to predict the behavior of the AV. In comparison, when the participants encountered the MV, the explicit information given by the driver's hand gesture after stopping the car helped them clearly understand the driving intentions and predict the driving behaviors, i.\,e., ``I am stopping'' and ``I will not depart before you cross the road.'' 

When the eHMI was used as a method of conveying explicit information from the AV to participants, 54.4\% of trials were evaluated as ``Strongly Agree'' and only 2.5\% of trials were evaluated as ``Disagree'' for Q1. This means that the participants' understanding of the driving intention for the AV w/ eHMI had a statistically significant improvement over that for the AV w/o eHMI (p$<$0.001).
Further, the participants in 42.5\% of trials strongly agreed that the behavior of the AV w/ eHMI was easy to predict, and only in 3.1\% of trials, they indicated disagreement; moreover, it was significantly easier to predict compared to predicting the behavior of the AV w/o eHMI (p$<$0.001). This illustrates that explicit information played a vital role in the pedestrian-AV interaction because there were no statistically significant differences between the evaluation results for the AV w/ eHMI and MV for Q1 and Q2.

After instructing the participants to enable them to understand the rationale of eHMI on AV correctly, they understood the driving intention and predicted the behavior of the AV more easily than before the instruction (Q1: p$<$0.05, Q2: p$<$0.01). In addition, the evaluation results between the MV and the AV w/ eHMI after instruction showed no statistical differences. The participants in 66.3\% of trials strongly agreed that the driving intention of the AV was easier to understand, which was greater than the ratio for the MV.
For Q2, 60.6\% of trials of the AV w/ eHMI after instruction were evaluated as ``Strongly Agree,'' which were more than the 56.3\% of trials of the MV evaluated as ``Strongly Agree''.
In addition, no trials were evaluated by Q2 as ``Disagree'' and ``Strongly Disagree'' when the participants encountered the AV w/ eHMI after instruction; however, 3.1\% of trials of the MV were evaluated by Q2 as ``Disagree''

The above results indicate that the instruction of the rationale of eHMI on AV to the pedestrians can not only help them easily understand the driving intention of the AV, but also aid them in more easily predicting the AV's behavior.

\subsection{Result of risk evaluation by Q3, Q4 and Q5}

Q3, Q4, and Q5 are related to the subjective risks of participants.
These subjective risks are perceived and evaluated based on the results of situational awareness, as shown in Fig.~\ref{fig:model}.
The dangerous feeling in the behavior of the car was evaluated by Q3. It is assumed to be based on the results of Q1 and Q2, that is, the participants would evaluate the danger of the car's behavior based on the perceived driving intention and predicted driving behavior. 
Further, the participants would adjust their trust in the car by comparing the result of the situational awareness with their experience. 
This result was evaluated by Q4. 
Moreover, Q5 evaluated the degrees of the feeling relief when crossing the road. It is assumed to be based on the results of Q3 and Q4. Pedestrians may cross the road relievedly when they do not feel that the behavior of the car is dangerous and they trust the car.

Results of Q3 in Fig.~\ref{fig:results} indicate that 26.3\% and 51.9\% of trials encountering the MV disagreed and strongly disagreed with the behaviors of the car being dangerous. 
Similarly, for Q5, 26.9\% and 64.4\% of trials encountering the MV agreed and strongly agreed that the participants were relieved when crossing the road. 
However, although 15.7\% (8.8\% + 6.9\%) of trials of the MV were evaluated as ``Agree'' and ``Strongly Agree'' in totality for Q3, only 2.5\% (2.5\% + 0.0\%) of trials of the MV were evaluated as not relieved by Q5.
This is attributed to the participants' trust in the driver; the participants may trust that the driver of MV would not cause harm to them even if they felt dangerous towards the MV.
This was illustrated by the results of Q4 which show that the 26.3\% and 60.0\% of trials of the MV were evaluated as ``Agree'' and ``Strongly Agree'', and only 2.5\% and 0.6\% of its trials were evaluated as ``Disagree'' and ``Strongly Disagree''. 

The results of Q3, Q4, and Q5 were significantly different when the participants encountered the AV w/o eHMI compared to those when the participants encountered the MV (p$<$0.001, respectively).
To compare with the evaluated results for the MV, the trials were evaluated as ``Strongly Disagree'' with danger in the behaviors of the AV w/o eHMI for Q3 being reduced to 21.3\%.
Further, the trials evaluated as ``Strongly Agree'' for trust in the AV w/o eHMI by Q4 reduced to 21.3\%.
These results were considered to be related to the results of Q1 and Q2.
That is, the participants increased dangerous feeling and decreased trust towards the AV because it was difficult to understand and predict the driving intention and behavior of the AV owing to the lack of explicit information.

The results of Q5 further indicate that the pedestrians strongly agreed that they relievedly crossed the road only in 24.4\% of trials when encountering the AV w/o eHMI.
In general, people are afraid under scenarios wherein they encounter things for which the behavior cannot be predicted.
Thus, the subjective risk of pedestrians would increase if the behaviors of AV were difficult to predict.

Compared to the AV w/o eHMI, the subjective evaluations of participants were significantly improved when the eHMI was used on the AV, as indicated by the results of Q3, Q4, and Q5 (p$<$0.001, respectively).
However, the subjective evaluation results of the AV w/ eHMI by Q4 and Q5 were not significantly better than those when the participants encountered the MV (p$<$0.01, respectively). For Q3, although there was no statistical difference between the subjective evaluation results for the MV and the AV w/ eHMI (p$>$0.05), the participants did not feel dangerous in 30.6\% of trials when they encountered the AV w/ eHMI compared to 51.9\% of trials that when they encountered the MV. The results of Q4 for the MV and the AV w/ eHMI showed statistical differences (p$<$0.01). Only 39.4\% of trials were evaluated as ``Strongly Agree'' for trust in the AV w/ eHMI, whereas it was 60.0\% of trials for the MV. This result suggests that using an eHMI on an AV can help improve the pedestrians' trust in an AV, but it cannot surpass their trust in an MV. 
Similarly, the results of Q5 showed that the trials of the AV w/ eHMI when the participants felt a sense of relief were lower than those for the trials of the MV (p$<$0.01) because they strongly agreed in 64.4\% of trials of the MV compared to 42.5\% of trials of the AV w/ eHMI.

After the instructing the rationale of eHMI on AV, the result of Q3 showed that the participants in 51.3\% and 38.1\% of trials disagreed and strongly disagreed that the behavior of the AV was dangerous.
Although these results of Q3 were not statistically different from the results of MV (p$>$0.05), the total ratio (51.3\% + 38.1\%) of ``Disagree'' and ``Strongly Disagree'' exceeded that of MV (26.3\% + 51.9\%). 
Further, only 3.1\% and 0.0\% of trials agreed and strongly agreed that the behavior of the AV w/ eHMI was dangerous after the instruction, which was less than the evaluations for the MV (8.8\% and 6.9\%). 
The results of Q4 indicated that the trust in the AV was no significant difference between encountering the MV and the AV w/ eHMI after instruction (p$>$0.05).
This result indicates that the participants trusted the MV and the AV w/ eHMI after instruction at the same level.
For Q5, when the participants were instructed, they felt a sense of relief in more trials than before the instruction (p$<$0.05). 
Besides, the results of Q5 show that the participants agreed and strongly agreed that they were relieved to cross the road in over 90.0\% of trials encountered the MV as same as the AV w/ eHMI after instruction (p$>$0.05).

The above results indicate that the subjective feelings and trust of pedestrians in an AV are at the same level as that for an MV when an eHMI is used on the AV after instructing the pedestrians.
Thus, after calibrating the correct mental model of pedestrians for the AV with the eHMI after instruction, the subjective feelings of the pedestrians improve when encountering the AV.

\subsection{Result of hesitation for decision making by Q6}
Q6 was used to evaluate the hesitant to make decisions when participants encountered the car, i.e., the difficulty of the decision-making.

The results of Q6 shown in Fig.~\ref{fig:results} indicate that the participants disagreed and strongly disagreed that they hesitated in a total of 67.6\% (33.8\% + 33.8\%) of trials when they encountered the MV. 
Further, in 14.4\% and 4.4\% of these trials, the participants agreed and strongly agreed that they hesitated in making a decision to cross the road. 

When the participants encountered the AV w/o eHMI, they hesitated in more trials than when they encountered the MV (p$<$0.001); in particular, they agreed that they hesitated in 42.5\% of trials when they encountered the AV w/o eHMI; however, they hesitated in only 14.4\% of trials when they encountered the MV.
This indicated that the absence of explicit information caused pedestrians to become hesitant to make decisions. 
Besides, the results of Q3, Q4, and Q5 are likely to affect the decision-making speed of the participants. 

With the eHMI being used on the AV, the ratio of trials wherein the participants agreed to have hesitated reduced to 18.1\%, and the ratio of trials wherein they disagreed and strongly disagreed to have hesitated increased to 65.0\% (35.0\% + 30.0\%) in totality.
This shows that the explicit information presented by the eHMI can help the participants quickly make a decision similar to the information provided by the hand gesture from the driver in the MV (p$>$0.05).

When the participants were instructed the rationale of the eHMI on AV, the ratio of trials wherein they disagreed and strongly disagreed to have hesitated further increased to 68.1\% (30.6\% + 37.5\%), which was almost the same as the 67.6\% (33.8\% + 33.8\%) of trials for the MV (p$>$0.05).

These results indicate that pedestrians can make decisions quickly once they are instructed to understand the rational of eHMI on AV correctly, i.\,e., once they can establish the mental model of the AV w/ eHMI correctly.

\section{CONCLUSION}
\vspace{-1mm}
In this paper, an interaction experiment was designed to allow participants to encounter an MV, AV w/o eHMI, AV w/ eHMI, and AV w/ eHMI after instruction under a road crossing scenario.

We found an obvious issue when the participants encountered the AV w/o eHMI.
The participants felt that it became more difficult to understand the driving intention and predict the driving behavior when they encountered the AV w/o eHMI after they had habituated to interacting with the MV. 
Further, their subjective feelings (i.\,e., dangerous feeling, trust in the AV, and feeling of relief) about the AV w/o eHMI deteriorated significantly, and they became hesitant to make a decision when they crossed the road. 
When the eHMI was used in the AV, it helped the participants understand the driving intention of AV and predict its driving behavior; further, their subjective feelings about the AV w/ eHMI and their hesitation towards decision-making improved. However, they still did not exceed the subjective evaluations of the MV.
After instructing the rationale of eHMI on AV to the pedestrians, it could help them correctly understand the driving intentions and predict the behavior of the AV, which could help improve their subjective feelings and decision-making.
Note that, those subjective evaluations reached the same standards as that for the MV.

In future works, we will further research the influence of instruction on the behaviors of participants by analyzing the recorded videos of them crossing the road.
We will refine the instruction and analyze the influence on pedestrians, such as providing instructions for cognition and risk evaluation. 
Finally, we hope to realize the standardization of the eHMI so that we can design a more accurate method of instruction.

\section*{ACKNOWLEDGMENT}
\vspace{-1mm}
This work was supported by JSPS KAKENHI Grant Numbers JP20K19846 and JP19K12080.

\bibliographystyle{ieeetr}
\bibliography{sample-base}

\end{document}